# EFFECT OF THREAD LEVEL PARALLELISM ON THE PERFORMANCE OF OPTIMUM ARCHITECTURE FOR EMBEDDED APPLICATIONS

Mehdi Alipour[1], Hojjat Taghdisi[2]

[1]Allameh Rafiei Highr Education Institute of Qazin, Iran
[2]Dept of Electrical, computer and IT engineering, Qazvin Islamic Azad University,Qazvin 34185-1416 Iran.

`mehdi.alipour@qiau.ac.ir, mehdi_10f@yahoo.com`

## ABSTRACT

*According to the increasing complexity of network application and internet traffic, network processor as a subset of embedded processors have to process more computation intensive tasks. By scaling down the feature size and emersion of chip multiprocessors (CMP) that are usually multi-thread processors, the performance requirements are somehow guaranteed. As multithread processors are the heir of uni-thread processors and there isn't any general design flow to design a multithread embedded processor, in this paper we perform a comprehensive design space exploration for an optimum uni-thread embedded processor based on the limited area and power budgets. Finally we run multiple threads on this architecture to find out the maximum thread level parallelism (TLP) based on performance per power and area optimum uni-thread architecture.*

## 1. INTRODUCTION

Embedded systems are designed to perform dedicated functions often with real-time computing constraints. While a PC or a general-purpose computer is designed to be flexible and can execute a wide range of applications. Embedded systems are used to control many devices in common use today [1]: more than 10 billion embedded processors have been sold in 2008 and more than 10.75 billion in 2009 [2].

In recent years embedded application and internet traffic have become heavier and sophisticated, so, future embedded processors will be encountered by more computation-intensive embedded applications and designing high performance processors is inevitable. By scaling down the feature size and emersion of chip multiprocessors (CMP) that are usually multi-thread processors, somehow the user's performance requirements are guaranteed. Recently in numerous researches, multi-thread processors are used to design a fast processor especially in network and embedded systems [3-7, 19].

In [20] a Markov model based on fine grain multithreading is implemented. Analytical Markov model is faster than simulation and has dispensable inaccuracy. In their method stalled threads are defined as states and transitions indicate the cache contention between threads [20]. Cache and register file are of the most important parts in designing multithread CMPs because the performance of a processor is severely related to cache access and also number of the registers. Cache memories are usually used to improve the performance and power consumption by bridging the gap between the speed and power consumption of the main memory and CPU. Therefore, the system performance and power consumption is severely related to the average memory access time and power consumption which makes cache as a major part in designing embedded processor architectures. In [3, 4] cache misses are introduced as a factor for reducing memory level parallelism between threads. Thread criticality prediction has been used in [5-8]. In these methods for achieving better performance, resources are given to the so called most critical threads which have higher L2 cache misses.

To improve packet-processing in network processors, [4, 8] have applied direct cache access (DCA) technique. In [4-9] processor architectures are based on simultaneous multithreading (SMT) and cache miss rate is used to evaluate the performance improvement. To find out the effect of cache access delay on performance, a comparison between multi-core and multi-thread processors has been performed in [12]. Likewise, victim cache is an approach to improve the performance of a multi-thread processor [5]. Most of the recent researches rely on comparing the multithread results with single-core single-thread processors. In the other word multi-thread processors are the heir of the single thread processors [6,7, 19]. Hence, evaluating the effective parameters such as cache and register file size is required for designing a multithread processor.

The first purpose of this paper is to study the effect of cache size on the performance because, embedded processors have to process computation and data intensive applications and  larger cache sizes will present better performance. Generally, one of the easiest ways to improve the performance of embedded and network processors is increasing the cache size [14-20], and [6,7] but this improvement, severely increase the occupied area and power consumption of the processor. So, it is necessary to find a cache size that creates the best tradeoff between performance, power, and area of the processor.  From other point of view, according to the performance per area parameter, higher performance in a specified area budget is one of the most important needs of a high performance embedded processor.

A negative point of the recent researches is that they don't have any constraints on the cache size. Because of the limited area budget in embedded processors, in this paper we will find the optimum size of L1 and L2 cache and also, because of the longer latency of bigger caches, best size of memory hierarchy in relation to this parameter has been explored. As mentioned above, another inseparable part in designing embedded processors is register file. Same as cache memory, size of this component has fundamental effect on the processor performance.

To improve the performance of an embedded processor, a large register file must be implemented. However, larger register files occupy more area and make a worse critical path [25]. Therefore, exploring the optimum size of the register file is the second purpose of this paper. The high importance of this issue is based on the fact that some parameters encourage designer to have a large register file. Generally embedded processors are implemented in multi-issue architectures and out of order (OOO) instruction execution that has renaming logic [23-25], [7], [13]. On the other hand, because register files are shared in multi-thread processors, making the common case fast, force the designer to have a larger register file [31]. In [22] effects of register file size in SMT processors have been studied. However, high budget for the number of registers has used.

The main contribution of our paper is to show the maximum number of threads that can be executed on a single-thread / single-core optimum architecture based on optimum performance per power of the cache and register file. We answer to these 2 important questions:

1- Is there any multi-thread architecture based on optimum single-thread architecture? (Area minimized architecture with limited power budget).
2- How much performance improvements can be reached by running multiple threads on optimum single-thread architecture. (Optimum multi-thread architectural guidelines based on optimum area and power budget of cache and register file).

## 2. SIMULATION ENVIRONMENT AND BENCHMARKS

For simulation, we used Multi2sim version 2.3.1[30], a super scalar multi-thread multi-core simulation platform which has 6 stages of pipeline for X86 instructions set architecture. It can run programs in multi-issue platform. We have changed and compiled the source code of the simulator on a 2.4 GHz, dual core processor with 4GB of RAM and 6MB of cache that run fedora 10 as an operating system.

Because embedded applications are so pervasive, homogenous applications cannot be a good choice for design space exploration (DSE). Hence we have done our DSE by heterogeneous applications from PacketBench [28] and MiBench[29]embedded benchmarks. Packetbench is a good platform to evaluate the workload characterization of network processors. Programs in this tool are categorized in 3 parts: *1-IP forwarding* which is corresponding to current internet standards. *2-packet classification* which is commonly used in firewalls and monitoring systems. *3- Encryption* which is a function that actually modifies the entire packet payload. Specific applications that we have used from each category are IPv4-Lctrie, Flow-Classification and IPSec.

 MiBench is a combination of six deferent categories. We have selected 3 of them. *1-Dijkstar* from network category. *2-susan (corners)* from automotive and industrial control category and *3-String-search* from office category. For a given source vertex (node) in the graph, the Dijkstra algorithm finds the path with the lowest cost (i.e. the shortest path) between that vertex and every other vertex. It can also be used for finding costs of shortest paths from a single vertex to a single destination vertex by stopping the algorithm once the shortest path to the destination vertex has been determined [29]. *Susan* is an image recognition package. It was developed for recognizing corners and edges in Magnetic Resonance Images of the brain [29] and *stringsearch* searches for given words in phrases using a case insensitive comparison algorithm.

## 3. EXPLORING CACHE ARCHITECTURE SPACE

### 3.1 Performance Analysis

This part is based on [26]. Authors of [26] did the exhaustive exploration of cache size for embedded application just based on performance and have introduced the cache size that produces lowest cycles for running an embedded application. Their research showed that there is a range for L1 and L2 cache for heterogeneous embedded applications. They showed that although performance is improved by increasing the cache size, however, over a threshold level performance is saturated and then decreased. Their range for cache size is too big so in this paper by considering another important parameter of embedded processor i.e. power or energy consumption, the range is reduced and just a few cache sizes for heterogeneous embedded applications are introduced.

Exploration of [20] reduced somehow 300 cache configurations to 36 configurations (6 sizes for L1 and 6 size for L2). Inthis paper, by considering both the dynamic and static power consumption of each configuration, we make more reduction on configurations of cache. We have calculated the best cache size for each application based on performance evaluation. Then the best performance for each application is calculated in the introduced size. From now we call this point of cache size the highest cache performance (HCP). HCP point produces lowest cycle simulation and HCP of all selected embedded applications are shown in fig.1.b in the right most column. Authors of [21] did somehow the same research based on [21] and [26], and they didn't consider power constraint of the cache which is very important in embedded processors.

## 3.2 Energy Analysis

To consider the power effects, we have used CACTI 5.0 [27], a tool from HP that is a platform for extracting parameters relevant to the cache size considering fabrication technology. Based on performance analysis of [26] there are 36 cache configurations for selected embedded applications. For calculating the power consumption of each configuration we have proposed the following model: Total energy that is consumed by a hardware module (here a cache) is calculated by adding total dynamic and static energy.

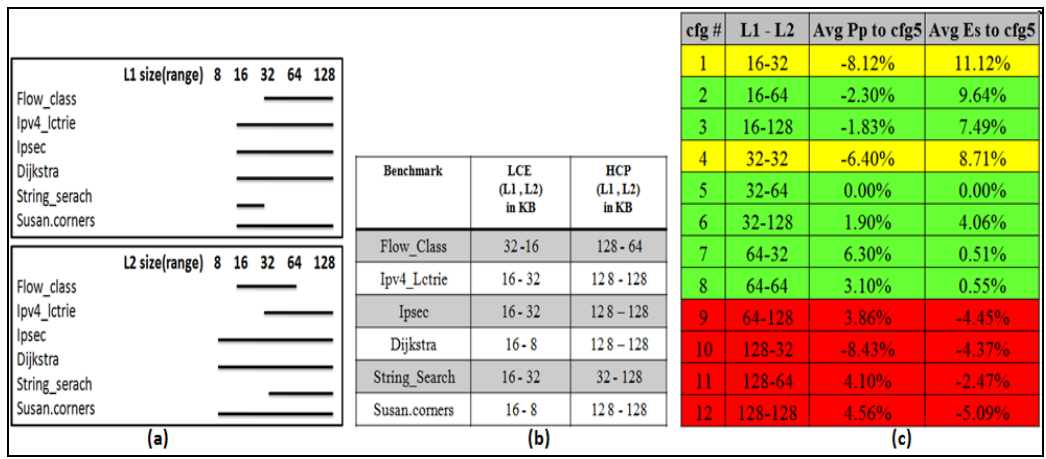

Figure1. a) Cache size overlapping. B) LCE and HCP points. C) Average performance penalty and energy saving

Dynamic energy is related to the supply voltage, module activity and output capacitance and clock frequency.

$$E_t = E_{td} + E_{ts} \,.$$

Where, $E_t$ is total energy dissipation, $E_{td}$ equals to the total dynamic energy and $E_{ts}$ is the total static energy (here in cache misses and the times that the cache is idle and there is no accesses to the specific level of cache). Any access to the cache is for reading or writing, so $E_{td}$ is affected by both reads and writes, so:

$$E_{td} = E_{dr} + E_{dw} \,.$$

Where $E_{dr}$ and $E_{dw}$ are dynamic read and write energy dissipations, respectively. In this paper we explore the cache memory in all levels including instruction cache level-1 ($L_1$), data cache level-1 (D$_1$) and unified cache level-2 (L$_2$). $E_{dr}$ is related to the number of reads (N$_{read}$) from all caches (number of read multiplied by dynamic energy of cache read), so:

$$Edr = [Nread(L1) * Edr(L1)] + [Nread(D1) * Edr(D1)] + [Nread(L2) * Edr(L2)] + [Nread(Main\_memory) * Edr(Main\_memory)].$$

And,

$$Edw = [Nwrite(L1) * Edw(L1)] + [Nwrite(D1) * Edw(D1)] + [Nwrite(L2) * Edw(L2) + [Nwrite(Main\_memory) * Edw(Main\_memory)].$$

Where, N$_{write}$ is the number of writes and for example E$_{dw}$ (D1) represents the dynamic energy of a write to D1. On the other hand, $E_{ts}$ is calculated from accumulating the consumed static energy ($E_s$) of all caches. In case of a cache miss, miss penalty which is related to the idle cache must be tolerated by the system. In this way, for a cache, miss penalty is considered as the cycles which are required for accessing the lower layer cache). Therefore:

$$Es = (Nmiss + idle\ cycles) * static\ energy\ per\ access * miss\ penalty\ (cycle).$$

Miss penalty is the cycle time consumed to access a next level cache and,

$$Ets = Es(L1) + Es(D1) + Es(L2).$$

Based on this model, each access consumes some energy considering the cache configuration and miss penalty. Although any access may lead to a miss or hit, however, any events cause some energy dissipation [34]. We have calculated the energy consumption of each cache configuration (dynamic and total separately) by using the proposed model, which considers the effect of all parameters i.e. number of cache misses/hits, access time of cache, cache level, type of accesses (read or write), and static/ dynamic energy on the energy dissipation of the cache.

Based on the energy analysis results, we introduced another point for cache configuration called lowest cache energy (LCE). LCEs are shown in fig.1.b in middle column and indicate the lowest energy consumption for each application. Results of fig.1.b show that for all applications, sizes of LCE is smaller than HCP so, LCE and HCP are the left and right margins of the cache size ranges, respectively, and they introduce a range for L1 and L2 considering both performance and energy consumption. To make a better sense for these ranges fig.1.a indicates the cache size overlapping of all applications in L1 and L2 ranges. Based on this figure, L1 (L2) range is from minimum L1 (L2) sizes for LCE column to maximum L1 (L2) sizes for HCP column. So in this way and based on fig1.a, L1 ranges from 8KB to 128KB and L2 ranges from 16KB to 128KB and we choose the cache sizes that have the most overlapping between all benchmarks. By using this ordinary and simple overlapping algorithm 36 cache configurations are reduced to 12.

These ranges specify an important point: any size for L1 and L2 out of this range is not recommended because the right side of these ranges leads to the maximum performance and the left side have the minimum power consumption for caches in selected embedded applications. Based on [20-26] the configuration of L1=32KB andL2 = 64KB are an applicable cache size for selected heterogeneous embedded applications. So we use it as the base of comparisons. Fig.1.c shows the average performance penalty (Pp column) of all 12 configurations related to this size which are labeled by cfg5 in the fig.1.c in the left most columns. As shown in fig.1.a, we can choose 16, 32, 64 and 128 KB for L1 and 32, 64, 128 KB for L2.

Table1. Performance penalty and energy saving of register file

| register# | Pp in avg | Es in avg |
|-----------|-----------|-----------|
| 48 | -1.87% | -0.64% |
| 56 | -1.11% | -0.62% |
| 64 | -0.46% | -0.22% |
| 72 | -0.15% | -0.03% |
| 80 | 0.00% | 0.00% |
| 88 | 0.01% | -0.04% |
| 96 | 0.01% | -0.10% |

So at most we have 12 points, (3 points for L1 and 4 points for L2) that are candidates to be the performance per energy optimum cache sizes useful for embedded processors.All of these 12 configurations are listed in fig.1.c. Since, performance per energy is one of the most important parameters in cache design for embedded processors, as indicated in the results; a cache size can be applicable only when satisfies these constraints. Based on fig.1.c cfg numbers from 1 to 4 have positive dynamic energy saving and negligible performance penalty. These sizes are the candidates for optimum and best cache configurations from performance and dynamic energy points of view.

Although cfg numbers 9, 11, and 12 create performance improvements but they consume higher dynamic energy and have no energy saving related to L1=32, L2 = 64 KB based on fig.1.c, so, they are not applicable cache sizes for selected embedded applications based on performance per power points of view. Therefore they are eliminated from search space and 12 configurations are reduced to 5 (cfg 1 to 5) it means 58% reduction in search space. From area point of view cfg 2 and cfg 3 are the best candidate for selected embedded application because they are smaller than others and also have positive energy saving and very low performance penalty. For running multiple threads we will use cfg 2 because it has 3% performance penalty that is tolerable by embedded applications [3-5].

## 4. EXPLORING OPTIMUM REGISTER FILE ARCHITECTURE

Performance evaluation of register file size have done by multi2sim but for power evaluation we have used McPAT [32] an integrated power, area, and timing modeling framework that supports comprehensive design space exploration for multi-core and many-core processor configurations ranging from 90nm to 22nm and beyond. McPAT can model both a reservation-station-model and a physical-register-file model based on real architectures.

To calculate the optimum size of register file, we have applied the parameters used for calculating best cache size, however, to find out just the effect of register file size on the performance, we used the best cache size (L1 and L2) concluded in the previous section for the cache size and run the simulator accordingly. Table1 shows the results of this part. In this table, 2 columns show the performance effect or performance penalty i.e. Pp and energy effect or energy saving i.e. Es of register file size. This table shows that although for all applications the best size of register file is 64 (in average) and above but in sizes near the half of this size, performance penalty is lower that 3%.

Also table1 shows that reducing the register file size always decrease the performance but sometimes, by doubling the register file size we don't have noticeable performance improvement.

So the first point that the highest performance is reached, is introduced as the best size for register file i.e. 80 registers. Based on power point of view the register file size=48 is the optimum size for selected embedded applications.

## 5. EXPLORING OPTIMUM MULTITHREAD ARCHITECTURE

In this part based on optimum sizes of cache and registerfile we introduce an optimum performance per power multi-thread architecture for selected embedded applications. It means Multi-threading upon uni-thread processor by running multiple thread on optimum single-thread/ single-core area minimized embedded processor. There are 3 type of multithreading that called *Interleaved multithreading* (IMT), *Blocked multithreading* (BMT) and *Simultaneous multithreading* (SMT) [33]. **IMT:** An instruction of another thread is fetched and fed into the execution pipeline at each processor cycle. **BMT:** The instructions of a thread are executed successively until an event occurs that may cause latency. This event induces a context switch. **SMT:** Instructions are simultaneously issued from multiple threads to the execution units of a superscalar processor. Thus, the wide superscalar instruction issue is combined with the multiple-context approach.

Researches show that SMT deliver highest performance improvements [3-7] so in this paper we used SMT architecture to access highest performance for embedded applications. The parameters we used for simulation to create multi-thread architecture are listed in table 2. Based on this table we used maximum sharing strategy to reach highest feasible performance improvements based on limited power and area budget. As multithreading creates some hardware redundancy, we used shared L1 cache and L2 cache, shared register file and shared all the parameters that can be shared between multiple threads. It means using minimum hardware for single-thread / single-core architectures that in average create just lower than 3% performance penalty and up to 7% energy saving for selected embedded applications.

The contribution is to answering this question: Is it feasible to run multiple threads on a single thread with limited area and power budget? If so, then what is the maximum number of threads that can be run on this architecture? And what is the maximum performance improvement? Results of this simulation are shown in figure 3. The results show that we can run up to 2 threads on a single-thread-single-core architecture and reach up to 31% performance improvement for susan corners benchmark and 17% performance improvements in average for selected embedded applications by running 2 threads on a single-thread processor with minimum area and power budget.

Table2. Processor parameters and sharing strategy.

| parameter name | (size), strategy |
|---|---|
| cache L1 | (16 KB), shared |
| cache L2 | (64 kB), shared |
| register file | (48), shared |
| fetch kind | shared |
| decode kind | shared |
| dispatch kind | shared |
| issue kind | shared |
| commit kind | shared |

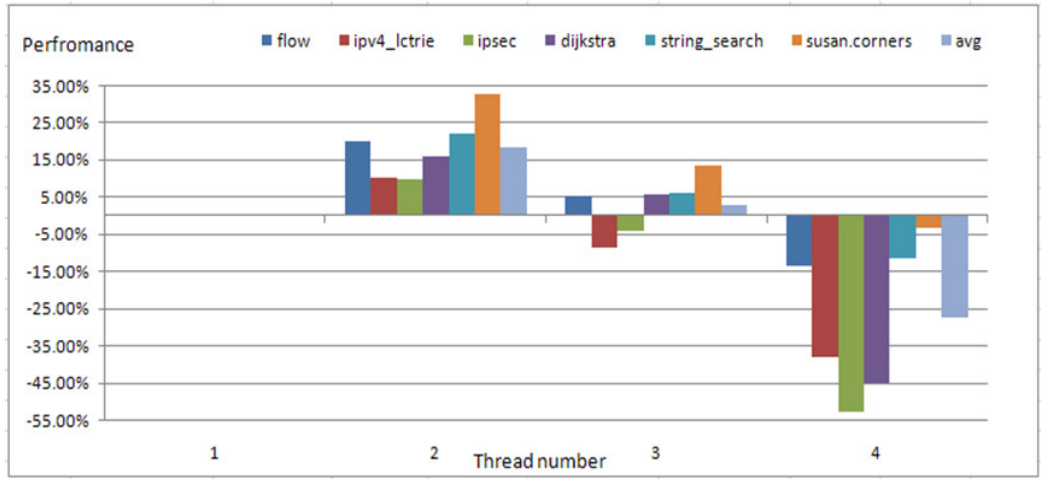

Figure3. Effect of multi-threading on the area minimized and optimum uni-thread architecture.

# 6. CONCLUSION

In this paper we performed an exhaustive design space exploration to find multi-thread architectural guideline for embedded application. Because multi-thread architectures are the heir of single-thread architectures we explored the optimum single-thread architecture based on cache and register file size as they are the most important components in embedded processors. We introduced optimum cache and register file size based on performance per power. We have executed multiple threads on a single-thread processor with limited area and power budget and results show that it is feasible to create multi-threading in a uni-thread architecture.

Because multi-threading causes to a hardware redundancy we explored the optimum size of cache and register file of the processor to have low area and just have negligible performance penalty (lower than 3%) and about 7% energy saving compared to the highest performance configuration. By this method, we created some room for hardware multi-threading. Our explorations show that running two threads on a single-thread processor with limited area and power budget, in average, leads to 17% performance improvements for selected representative embedded applications.